\documentclass[preprint,12pt]{elsarticle}

\usepackage{amssymb}

\usepackage{amsthm}

\usepackage{amsmath}
\DeclareMathOperator{\tr}{tr}

\DeclareMathOperator{\supp}{supp}
\newcommand{\Slash}[1]{{\ooalign{\hfil/\hfil\crcr$#1$}}}
\numberwithin{equation}{section}

\usepackage{url}

\begin{document}

\begin{frontmatter}



\title{%
Remark on the energy-momentum tensor in the lattice
formulation of 4D $\mathcal{N}=1$ SYM}


\author{Hiroshi Suzuki}
\ead{hsuzuki@riken.jp}
\address{%
Theoretical Research Division, RIKEN Nishina Center, Wako 2-1, Saitama
351-0198, Japan}

\begin{abstract}
In a recent paper, arXiv:1209.2473~\cite{Suzuki:2012gi}, we presented a
possible definition of the energy-momentum tensor in the lattice formulation
of the four-dimensional $\mathcal{N}=1$ supersymmetric Yang--Mills theory, that
is conserved in the quantum continuum limit. In the present Letter, we propose
a quite similar but somewhat different definition of the energy-momentum tensor
(that is also conserved in the continuum limit) which is superior in several
aspects: In the continuum limit, the origin of the energy automatically becomes
consistent with the supersymmetry and the number of renormalization constants
that require a (non-perturbative) determination is reduced to two from four,
the number of renormalization constants appearing in the construction
in~Ref.~\cite{Suzuki:2012gi}.
\end{abstract}

\end{frontmatter}


\section{Introduction}
\label{sec:1}
Although the energy-momentum tensor is a very fundamental observable in field
theory, it is not straightforward to define the energy-momentum tensor in the
lattice field theory, because the spacetime lattice explicitly breaks
translational and rotational symmetries. For four-dimensional lattice gauge
theories containing fermions, a strategy to construct an energy-momentum
tensor, that satisfies the conservation law in the quantum continuum limit, has
been given in~Ref.~\cite{Caracciolo:1989pt}. In quantum field theory, a
symmetry is generally expressed by corresponding Ward--Takahashi (WT) relations
and the conservation law is merely a special case of WT relations that holds
only when the Noether current stays away from other operators. Nevertheless, as
demonstrated in~Ref.~\cite{Caracciolo:1989pt} (and probably as can be proven
generally), any lattice energy-momentum tensor, that is conserved in the
continuum limit, is expected to reproduce all WT relations associated with the
translational invariance for \emph{elementary fields} in the continuum
limit.\footnote{On the other hand, at the current moment there is no analysis
on how one can construct a lattice energy-momentum tensor that generates
correctly-normalized translations on \emph{composite operators}. The
complication arises because one has to classify the operator mixing occurring
when the energy-momentum tensor and composite operators coincide in position
space.} This shows the fundamental importance of the conservation law in the
continuum limit for a lattice energy-momentum tensor.

The present Letter is an extension of our recent paper~\cite{Suzuki:2012gi}
concerning the energy-momentum tensor in the lattice formulation of the
four-dimensional $\mathcal{N}=1$ supersymmetric Yang--Mills theory (4D
$\mathcal{N}=1$ SYM). In~Ref.~\cite{Suzuki:2012gi}, we proposed a possible
lattice energy-momentum tensor by mimicking the structure of the
Ferrara--Zumino (FZ) supermultiplet~\cite{Ferrara:1974pz}. That is, we defined
a lattice energy-momentum tensor by a renormalized, modified supersymmetry
(SUSY) transformation of a renormalized SUSY current on the lattice. Then,
assuming the locality and the hypercubic symmetry of the lattice formulation
and that the bare gluino mass is tuned so that the SUSY current is
conserved~\cite{Curci:1986sm,Donini:1997hh}, the energy-momentum tensor was
shown to be conserved in the quantum continuum limit; as noted above, this is a
minimal and fundamental requirement on the energy-momentum tensor. This lattice
energy-momentum tensor can be a basic tool to compute physical quantities
related to the energy-momentum tensor, such as the viscosity.

Although the general strategy to construct a conserved lattice energy-momentum
tensor in~Ref.~\cite{Caracciolo:1989pt} is applicable also to the lattice
formulation of 4D $\mathcal{N}=1$ SYM, our method that is based on the
$\mathcal{N}=1$ SUSY in the target theory is interesting, because the direct
imposition of the conservation law requires the (non-perturbative)
determination of at least six renormalization
constants~\cite{Caracciolo:1989pt}, while the method
in~Ref.~\cite{Suzuki:2012gi} contains only four (or three if one does not care
about the ambiguity of the zero-point energy) unknown renormalization
constants; see below.

In the present Letter, as a possible alternative of the definition
in~Ref.~\cite{Suzuki:2012gi}, we propose a quite similar but somewhat different
definition of a lattice energy-momentum tensor for 4D $\mathcal{N}=1$ SYM; this
energy-momentum tensor is also conserved in the continuum limit. This new
definition is superior in several aspects compared with the one
in~Ref.~\cite{Suzuki:2012gi}: In the continuum limit, the origin of the energy
automatically becomes consistent with SUSY and the number of renormalization
constants that require a (non-perturbative) determination is reduced to two
from four, the number of renormalization constants appearing in the
construction in~Ref.~\cite{Suzuki:2012gi}. We follow the notational convention
of~Ref.~\cite{Suzuki:2012gi}.\footnote{Vector indices $\mu$, $\nu$, \dots, run
over $0$, $1$, $2$, $3$. $\epsilon_{\mu\nu\rho\sigma}$~denotes the totally
anti-symmetric tensor and~$\epsilon_{0123}=-1$. All gamma matrices are
hermitian and obey $\{\gamma_\mu,\gamma_\nu\}=2\delta_{\mu\nu}$. We define
$\gamma_5\equiv-\gamma_0\gamma_1\gamma_2\gamma_3$
and~$\sigma_{\mu\nu}\equiv[\gamma_\mu,\gamma_\nu]/2$. The charge conjugation
matrix~$C$ satisfies, $C^{-1}\gamma_\mu C=-\gamma_\mu^T$,
$C^{-1}\sigma_{\mu\nu}C=-\sigma_{\mu\nu}^T$, $C^{-1}\gamma_5C=\gamma_5^T$
and~$C^T=-C$. The generator of the gauge group $SU(N_c)$, $T^a$, is normalized
as $\tr(T^aT^b)=(1/2)\delta^{ab}$. $g$~is the bare gauge coupling constant.
$x$, $y$, $z$, \dots\ denote lattice points and $a$~is the lattice spacing;
$\Hat{\mu}$ is the unit vector in the $\mu$-direction. $U_\mu(x)\in SU(N_c)$
denotes the conventional link variable and $\psi(x)\in su(N_c)$ is the gluino
field and~$\Bar{\psi}(x)\equiv\psi^T(x)(-C^{-1})$. The symmetric difference
operator $\partial_\mu^S$ is defined by
\begin{equation}
   \partial_\mu^Sf(x)\equiv\frac{1}{2a}
   \left[f(x+a\Hat{\mu})-f(x-a\Hat{\mu})\right].
\label{eq:(1.1)}
\end{equation}}

\section{A new definition of the energy-momentum tensor on the lattice}
As Ref.~\cite{Suzuki:2012gi}, our starting point for the construction of a
lattice energy-momentum tensor is a renormalized SUSY WT relation on the
lattice:
\begin{equation}
   \left\langle
   \partial_\mu^S\mathcal{S}_\mu(x)
   \mathcal{O}\right\rangle
   =\left\langle
   \mathcal{Z}
   \left[-\frac{1}{a^4}\frac{\partial}{\partial\Bar{\xi}(x)}\Delta_\xi
   +a\mathcal{E}(x)\right]
   \mathcal{O}\right\rangle.
\label{eq:(2.1)}
\end{equation}
Throughout the present Letter, we assume that the composite operator denoted
by~$\mathcal{O}$ is gauge invariant and finite, i.e., it is already
appropriately renormalized. In the left-hand of~Eq.~\eqref{eq:(2.1)},
$\mathcal{S}_\mu(x)$ is a renormalized Noether current associated with SUSY
(the renormalized SUSY current),
\begin{equation}
   \mathcal{S}_\mu(x)
   \equiv\mathcal{Z}
   \left[\mathcal{Z}_SS_\mu(x)+\mathcal{Z}_TT_\mu(x)\right],
\label{eq:(2.2)}
\end{equation}
where $\mathcal{Z}$, $\mathcal{Z}_S$ and~$\mathcal{Z}_T$ are renormalization
constants\footnote{The multiplicative renormalization constant~$\mathcal{Z}$ is
chosen so that the operator $\mathcal{S}_\mu(x)$ has a finite correlation
function with any renormalized operator, when the point~$x$ is far apart from
the support of that operator by a finite physical distance. $\mathcal{Z}$ is at
most logarithmically divergent by a dimensional reason.}
and lattice operators $S_\mu(x)$ and~$T_\mu(x)$ are defined by
\begin{align}
   S_\mu(x)&\equiv-\sigma_{\rho\sigma}\gamma_\mu
   \tr\left\{\psi(x)\left[F_{\rho\sigma}\right]^L(x)\right\},
\notag\\
   T_\mu(x)&\equiv2\gamma_\nu
   \tr\left\{\psi(x)\left[F_{\mu\nu}\right]^L(x)\right\}.
\label{eq:(2.3)}
\end{align}
Here and in what follows, $[F_{\mu\nu}]^L(x)$ denotes a lattice transcription of
the field strength,
\begin{equation}
   \left[F_{\mu\nu}\right]^L(x)\equiv2\tr\left[P_{\mu\nu}(x)T^a\right]T^a,
\label{eq:(2.4)}
\end{equation}
defined from the clover plaquette $P_{\mu\nu}(x)$,
\begin{equation}
   P_{\mu\nu}(x)\equiv\frac{1}{4}\sum_{i=1}^4\frac{1}{2ia^2g}
   \left[U_{i\mu\nu}(x)-U_{i\mu\nu}^\dagger(x)\right],
\label{eq:(2.5)}
\end{equation}
where
\begin{align}
   U_{1\mu\nu}(x)
   &\equiv U_\mu(x)U_\nu(x+a\hat\mu)U_\mu^\dagger(x+a\hat\nu)U_\nu^\dagger(x),
\notag\\
   U_{2\mu\nu}(x)
   &\equiv U_\nu(x)U_\mu^\dagger(x-a\hat\mu+a\hat\nu)
   U_\nu^\dagger(x-a\hat\mu)U_\mu(x-a\hat\mu),
\notag\\
   U_{3\mu\nu}(x)
   &\equiv U_\mu^\dagger(x-a\hat\mu)U_\nu^\dagger(x-a\hat\mu-a\hat\nu)
   U_\mu(x-a\hat\mu-a\hat\nu)U_\nu(x-a\hat\nu),
\notag\\
   U_{4\mu\nu}(x)
   &\equiv U_\nu^\dagger(x-a\hat\nu)U_\mu(x-a\hat\nu)
   U_\nu(x+a\hat\mu-a\hat\nu)U_\mu^\dagger(x).
\label{eq:(2.6)}
\end{align}
In the right-hand side of~Eq.~\eqref{eq:(2.1)}, $\Delta_\xi$ is a modified SUSY
transformation on lattice variables with the localized transformation
parameter~$\xi(x)$,
\begin{equation}
   \Delta_\xi\equiv
   \delta_\xi+\mathcal{Z}_{\text{EOM}}\delta_{F\xi},
\label{eq:(2.7)}
\end{equation}
which depends on another renormalization
constant~$\mathcal{Z}_{\text{EOM}}$~\cite{Suzuki:2012gi}; the localized
transformations $\delta_\xi$ and~$\delta_{F\xi}$ are defined by
($\Bar{\xi}(x)\equiv\xi^T(x)(-C^{-1})$)
\begin{align}
   \delta_\xi U_\mu(x)&\equiv iag\frac{1}{2}
   \left[
   \Bar{\xi}(x)\gamma_\mu\psi(x)U_\mu(x)
   +\Bar{\xi}(x+a\hat\mu)\gamma_\mu U_\mu(x)\psi(x+a\hat\mu)
   \right],
\notag\\
   \delta_\xi U_\mu^\dagger(x)&\equiv-iag\frac{1}{2}
   \left[
   \Bar{\xi}(x)\gamma_\mu U_\mu^\dagger(x)\psi(x)
   +\Bar{\xi}(x+a\hat\mu)\gamma_\mu\psi(x+a\hat\mu)U_\mu^\dagger(x)
   \right],
\notag\\
   \delta_\xi\psi(x)&\equiv-\frac{1}{2}\sigma_{\mu\nu}\xi(x)
   \left[F_{\mu\nu}\right]^L(x),\qquad
   \delta_\xi\Bar{\psi}(x)
   =\frac{1}{2}\Bar{\xi}(x)\sigma_{\mu\nu}\left[F_{\mu\nu}\right]^L(x),
\label{eq:(2.8)}
\end{align}
and
\begin{equation}
   \delta_{F\xi}U_\mu(x)=0,\qquad
   \delta_{F\xi}\psi(x)=\delta_\xi\psi(x),\qquad
   \delta_{F\xi}\Bar{\psi}(x)=\delta_\xi\Bar{\psi}(x).
\label{eq:(2.9)}
\end{equation}
Finally, $\mathcal{E}(x)$ in~Eq.~\eqref{eq:(2.1)} is a dimension~$11/2$
operator that is given by a linear combination of renormalized operators with
logarithmically divergent coefficients.

The derivation of the renormalized SUSY WT relation~\eqref{eq:(2.1)} is
somewhat too lengthy to be reproduced here; we refer the interested reader
to~Ref.~\cite{Suzuki:2012gi} and references cited therein, especially for the
origin of various renormalization constants. Here, we simply note that
Eq.~\eqref{eq:(2.1)} reduces to the conservation law of the renormalized SUSY
current~$\mathcal{S}_\mu(x)$ in the continuum limit, when the point~$x$ stays
away from the support of the operator~$\mathcal{O}$ by a finite physical
distance (we express this situation
by~$x\leftrightsquigarrow\supp(\mathcal{O})$),
\begin{equation}
   \left\langle
   \partial_\mu^S\mathcal{S}_\mu(x)
   \mathcal{O}\right\rangle
   \xrightarrow{a\to0}0,\qquad
   \text{for $x\leftrightsquigarrow\supp(\mathcal{O})$}.
\label{eq:(2.10)}
\end{equation}
This follows because in the right-hand side of~Eq.~\eqref{eq:(2.1)}, the
$\Bar{\xi}(x)$ derivative vanishes and the dimension~$11/2$
operator~$\mathcal{E}(x)$ does not produce an $O(1/a)$ linear-divergence that
could compensate the factor~$a$ when~$x\leftrightsquigarrow\supp(\mathcal{O})$.
In deriving Eq.~\eqref{eq:(2.1)}, we assumed that the the bare gluino mass~$M$
is tuned to the supersymmetric point~\cite{Curci:1986sm,Donini:1997hh,%
Taniguchi:1999fc,Farchioni:2001wx,Suzuki:2012pc} and that there is no exotic
SUSY anomaly of the form of a three-fermion operator~\cite{Farchioni:2001wx,%
Suzuki:2012pc}. The relation~\eqref{eq:(2.10)} can be regarded as the
restoration of SUSY (that is broken by the lattice regularization) in the
continuum limit.

In~Ref.~\cite{Suzuki:2012gi}, a symmetric energy-momentum tensor on the lattice
was defined by,
\begin{equation}
   \mathcal{T}_{\mu\nu}(x)
   \equiv\frac{1}{2}\left[\varTheta_{\mu\nu}(x)+\varTheta_{\nu\mu}(x)\right]
   -c\delta_{\mu\nu}
   \tr\left[\Bar{\psi}(x)(D+M)\psi(x)\right],
\label{eq:(2.11)}
\end{equation}
where $D$ denotes the lattice Dirac operator and\footnote{The subscripts
$\alpha$ and~$\beta$ refer to the spinor index.}
\begin{equation}
   \varTheta_{\mu\nu}(x)
   \equiv\frac{1}{8}(\gamma_\nu)_{\beta\alpha}
   \frac{\partial}{\partial\xi_\beta}
   \left[\mathcal{Z}\Bar{\Delta}_\xi\mathcal{S}_\mu(x)\right]_\alpha,
\label{eq:(2.12)}
\end{equation}
and $\Bar{\Delta}_\xi$ is a global modified SUSY transformation on lattice
variables, that is obtained by setting the local parameter
constant, $\xi(x)\to\xi$, in~Eq.~\eqref{eq:(2.7)}. $c$~in~Eq.~\eqref{eq:(2.11)}
is a constant to be fixed, although it does not affect the conservation
of~$\mathcal{T}_{\mu\nu}(x)$. Using the SUSY WT relation~\eqref{eq:(2.1)}, it
can be shown that the energy-momentum tensor~\eqref{eq:(2.11)} is conserved in
the continuum limit~\cite{Suzuki:2012gi}. The definition
through~Eqs.~\eqref{eq:(2.11)} and~\eqref{eq:(2.12)} was suggested by the
structure of the FZ supermultiplet~\cite{Ferrara:1974pz} that the SUSY
transformation of the SUSY current is basically the energy-momentum tensor.

Now, our new definition of a lattice energy-momentum tensor proceeds as
follows: By using the renormalized SUSY current~\eqref{eq:(2.2)}, we first
define the quantity,
\begin{equation}
   \varTheta_{\mu\nu}(x;\mathcal{D}_x)
   \equiv-\frac{1}{8}\left(C^{-1}\gamma_\nu\right)_{\alpha\beta}
   a^4\sum_{y\in\mathcal{D}_x}
   \left[\partial_\rho^S\mathcal{S}_\rho(y)\right]_\alpha
   \left[\mathcal{S}_\mu(x)\right]_\beta,
\label{eq:(2.13)}
\end{equation}
where $\mathcal{D}_x$ is a hypercubic region on the lattice that contains the
SUSY current~$\mathcal{S}_\mu(x)$ entirely; the point~$x$ is taken as the
center of the region~$\mathcal{D}_x$ so that $\mathcal{D}_x$ is invariant
under the hypercubic rotation around~$x$. Moreover, the size of the
region~$\mathcal{D}_x$ must be ``macroscopic'', i.e., it must be finite in the
physical unit. The definition of $\varTheta_{\mu\nu}(x;\mathcal{D}_x)$ thus
depends on the choice of the region~$\mathcal{D}_x$ as its argument indicates.
From this $\varTheta_{\mu\nu}(x;\mathcal{D}_x)$, we define a symmetric
energy-momentum tensor on the lattice, simply by symmetrizing it with respect
to the indices:
\begin{equation}
   \mathcal{T}_{\mu\nu}(x;\mathcal{D}_x)
   \equiv\frac{1}{2}\left[
   \varTheta_{\mu\nu}(x;\mathcal{D}_x)
   +\varTheta_{\nu\mu}(x;\mathcal{D}_x)\right].
\label{eq:(2.14)}
\end{equation}

The idea behind the definition in~Eqs.~\eqref{eq:(2.13)} and~\eqref{eq:(2.14)}
is as follows: In the continuum theory, at least formally, the integral of the
total divergence of the SUSY current in the continuum
theory~$\Breve{S}_\rho(y)$,
$\int_{\mathcal{D}_x}d^4y\,\partial_\rho\Breve{S}_\rho(y)$, where the
region~$\mathcal{D}_x$ contains an operator at the point~$x$, generates the
SUSY transformation,
\begin{equation}
   -\int d^4y\,\frac{\delta}{\delta\Bar{\xi}(y)}\delta_\xi
   =-\frac{\partial}{\partial\Bar{\xi}}\Bar{\delta}_\xi,
\label{eq:(2.15)}
\end{equation}
on the operator ($\delta_\xi$ and~$\Bar{\delta}_\xi$ are localized and global
SUSY transformations, respectively). In the classical continuum theory, on the
other hand, the energy-momentum tensor~$\Breve{T}_{\mu\nu}(x)$ is given by the
SUSY transformation of the SUSY current~\cite{Ferrara:1974pz}
as (see~Ref.~\cite{Suzuki:2012gi}),
\begin{align}
   \Breve{\varTheta}_{\mu\nu}(x)
   &\equiv\frac{1}{8}\left(C^{-1}\gamma_\nu\right)_{\alpha\beta}
   \frac{\partial}{\partial\Bar{\xi}_\alpha}
   \left[\Bar{\delta}_\xi\Breve{S}_\mu(x)\right]_\beta,
\label{eq:(2.16)}
\\
   \Breve{T}_{\mu\nu}(x)
   &=\frac{1}{2}\left[
   \Breve{\varTheta}_{\mu\nu}(x)
   +\Breve{\varTheta}_{\nu\mu}(x)\right]
   -\frac{3}{4}\delta_{\mu\nu}
   \tr\left[\Bar{\psi}(x)\Slash{D}\psi(x)\right],
\label{eq:(2.17)}
\end{align}
where $\Slash{D}$ denotes the Dirac operator. Thus one sees that the
definition~\eqref{eq:(2.13)} is a lattice transcription of the relation
expected in the continuum theory,\footnote{On the other hand, in transcribing
Eq.~\eqref{eq:(2.17)} to the lattice theory~\eqref{eq:(2.14)}, we discarded the
last term $-(3/4)\delta_{\mu\nu}\tr[\Bar{\psi}(x)\Slash{D}\psi(x)]$. In quantum
theory, this term just acts as the zero-point energy
(see~Ref.~\cite{Suzuki:2012gi}) and we will see below that the simple
prescription~\eqref{eq:(2.14)} gives rise to the zero-point energy that is
consistent with SUSY.}
\begin{equation}
   \Breve{\varTheta}_{\mu\nu}(x)
   =-\frac{1}{8}(C^{-1}\gamma_\nu)_{\alpha\beta}
   \int_{\mathcal{D}_x}d^4y\,\left[\partial_\rho\Breve{S}_\rho(y)\right]_\alpha
   \left[\Breve{S}_\mu(x)\right]_\beta.
\label{eq:(2.18)}
\end{equation}
In the classical continuum theory, the right-hand side of Eq.~\eqref{eq:(2.18)}
is independent of the choice of the region~$\mathcal{D}_x$ because of the
current conservation. In the lattice theory, however, this property is lost
because the conservation law of the SUSY current is broken by $O(a)$ terms.
That is, the dependence on~$\mathcal{D}_x$ in~Eqs.~\eqref{eq:(2.13)}
and~\eqref{eq:(2.14)} is an $O(a)$ lattice artifact and the physics in the
continuum limit should not depend on the choice of~the
region~$\mathcal{D}_x$.\footnote{By an argument similar to the one in what
follows, it is easy to see that the difference
in~$\mathcal{T}_{\mu\nu}(x;\mathcal{D}_x)$ due to different choices
of~$\mathcal{D}_x$ vanishes in the continuum limit, at least when the
energy-momentum tensor and other renormalized operators are separated to each
other by finite physical distances. This shows that, in particular, the
expectation value of~$\mathcal{T}_{\mu\nu}(x;\mathcal{D}_x)$ with respect to
physical states becomes independent of the choice of~$\mathcal{D}_x$ in the
continuum limit.}

We note that the energy-momentum tensor~\eqref{eq:(2.14)} is manifestly finite,
because the operator~$\sum_{y\in\mathcal{D}_x}\partial_\rho^S\mathcal{S}_\rho(y)$
in~Eq.~\eqref{eq:(2.13)}, being the sum of the total divergence, does not have
any overlap with the operator~$\mathcal{S}_\mu(x)$; Eq.~\eqref{eq:(2.13)} is
thus the sum of products of renormalized operators at points separated by
finite physical distances.

Let us show that the lattice energy-momentum
tensor~$\mathcal{T}_{\mu\nu}(x;\mathcal{D}_x)$~\eqref{eq:(2.14)} is conserved in
the continuum limit. For this, we first show the conservation
of~$\varTheta_{\mu\nu}(x;\mathcal{D}_x)$~\eqref{eq:(2.13)}: Let~$2R$ be the size
of~$\mathcal{D}_x$,
\begin{equation}
   \mathcal{D}_x\equiv\left\{y\in L^4\mid
   \text{$x_\mu-R\leq y_\mu\leq x_\mu+R$ for all~$\mu$}\right\},
\label{eq:(2.19)}
\end{equation}
where $L^4$ denotes the whole lattice of the size~$L^4$, and define a
three-dimensional cubic region orthogonal to the $\mu$-direction as
\begin{equation}
   \mathcal{C}_x^{(\mu)}(z_\mu)\equiv\left\{y\in L^4\mid
   \text{$x_\nu-R\leq y_\nu\leq x_\nu+R$ for $\nu\neq \mu$ and
   $y_\mu=z_\mu$}\right\}.
\label{eq:(2.20)}
\end{equation}
Then, from the definition~\eqref{eq:(2.13)} and the SUSY WT
relation~\eqref{eq:(2.1)}, we have
\begin{align}
   &\left\langle
   \partial_\mu^S\varTheta_{\mu\nu}(x;\mathcal{D}_x)
   \mathcal{O}
   \right\rangle
\notag\\
   &=\frac{1}{8}\left(C^{-1}\gamma_\nu\right)_{\alpha\beta}
   a^4\sum_{y\in\mathcal{D}_x}
   \left\langle
   \mathcal{Z}
   \left[-\frac{1}{a^4}\frac{\partial}{\partial\Bar{\xi}(x)}\Delta_\xi
   +a\mathcal{E}(x)\right]_\beta
   \left[\partial_\rho^S\mathcal{S}_\rho(y)\right]_\alpha
   \mathcal{O}
   \right\rangle
\notag\\
   &\qquad{}
   -\frac{1}{8}\left(C^{-1}\gamma_\nu\right)_{\alpha\beta}
   \sum_\mu\frac{1}{2}\left[a^3\sum_{y\in\mathcal{C}_x^{(\mu)}(x_\mu+R+a)}
   -a^3\sum_{y\in\mathcal{C}_x^{(\mu)}(x_\mu-R)}\right]
\notag\\
   &\qquad\qquad\qquad\qquad{}
   \times\left\langle
   \mathcal{Z}
   \left[-\frac{1}{a^4}\frac{\partial}{\partial\Bar{\xi}(y)}\Delta_\xi
   +a\mathcal{E}(y)\right]_\alpha
   \left[\mathcal{S}_\mu(x+a\Hat{\mu})\right]_\beta
   \mathcal{O}
   \right\rangle
\notag\\
   &\qquad{}
   -\frac{1}{8}\left(C^{-1}\gamma_\nu\right)_{\alpha\beta}
   \sum_\mu\frac{1}{2}\left[a^3\sum_{y\in\mathcal{C}_x^{(\mu)}(x_\mu+R)}
   -a^3\sum_{y\in\mathcal{C}_x^{(\mu)}(x_\mu-R-a)}\right]
\notag\\
   &\qquad\qquad\qquad\qquad{}
   \times\left\langle
   \mathcal{Z}
   \left[-\frac{1}{a^4}\frac{\partial}{\partial\Bar{\xi}(y)}\Delta_\xi
   +a\mathcal{E}(y)\right]_\alpha
   \left[\mathcal{S}_\mu(x-a\Hat{\mu})\right]_\beta
   \mathcal{O}
   \right\rangle.
\label{eq:(2.21)}
\end{align}

Suppose now that the point~$x$ stays away from the support of the
operator~$\mathcal{O}$ by a finite physical distance,
$x\leftrightsquigarrow\supp(\mathcal{O})$, and the region~$\mathcal{D}_x$ has
been chosen such that~$\mathcal{D}_x\cap\supp(\mathcal{O})=\varnothing$. In
this situation, Eq.~\eqref{eq:(2.21)} reduces to
\begin{align}
   &\left\langle
   \partial_\mu^S\varTheta_{\mu\nu}(x;\mathcal{D}_x)
   \mathcal{O}
   \right\rangle
\notag\\
   &=\frac{1}{8}\left(C^{-1}\gamma_\nu\right)_{\alpha\beta}
   \left\langle
   \mathcal{Z}
   \left[a\mathcal{E}(x)\right]_\beta
   \left[a^4\sum_{y\in\mathcal{D}_x}
   \partial_\rho^S\mathcal{S}_\rho(y)\right]_\alpha
   \mathcal{O}
   \right\rangle
\notag\\
   &\qquad{}
   -\frac{1}{8}\left(C^{-1}\gamma_\nu\right)_{\alpha\beta}
   \sum_\mu\frac{1}{2}\left[a^3\sum_{y\in\mathcal{C}_x^{(\mu)}(x_\mu+R+a)}
   -a^3\sum_{y\in\mathcal{C}_x^{(\mu)}(x_\mu-R)}\right]
\notag\\
   &\qquad\qquad\qquad\qquad{}
   \times\left\langle
   \mathcal{Z}
   \left[a\mathcal{E}(y)\right]_\alpha
   \left[\mathcal{S}_\mu(x+a\Hat{\mu})\right]_\beta
   \mathcal{O}
   \right\rangle
\notag\\
   &\qquad{}
   -\frac{1}{8}\left(C^{-1}\gamma_\nu\right)_{\alpha\beta}
   \sum_\mu\frac{1}{2}\left[a^3\sum_{y\in\mathcal{C}_x^{(\mu)}(x_\mu+R)}
   -a^3\sum_{y\in\mathcal{C}_x^{(\mu)}(x_\mu-R-a)}\right]
\notag\\
   &\qquad\qquad\qquad\qquad{}
   \times\left\langle
   \mathcal{Z}
   \left[a\mathcal{E}(y)\right]_\alpha
   \left[\mathcal{S}_\mu(x-a\Hat{\mu})\right]_\beta
   \mathcal{O}
   \right\rangle.
\label{eq:(2.22)}
\end{align}
Now noting that the
combination~$\sum_{y\in\mathcal{D}_x}\partial_\rho^S\mathcal{S}_\rho(y)$ does not
have any overlap with the point~$x$, we see that Eq.~\eqref{eq:(2.22)} is the
sum of correlation functions of renormalized operators with no mutual overlap
with an overall factor of~$a$ (in front of the operator~$\mathcal{E}(x)$).
Thus, Eq.~\eqref{eq:(2.22)} vanishes in the $a\to0$ limit and
$\varTheta_{\mu\nu}(x;\mathcal{D}_x)$ is conserved in the continuum limit:
\begin{equation}
   \left\langle
   \partial_\mu^S\varTheta_{\mu\nu}(x;\mathcal{D}_x)\mathcal{O}
   \right\rangle
   \xrightarrow{a\to0}0,
   \qquad\text{for $x\leftrightsquigarrow\supp(\mathcal{O})$}.
\label{eq:(2.23)}
\end{equation}

Next, we consider the anti-symmetric part
of~$\varTheta_{\mu\nu}(x;\mathcal{D}_x)$,
\begin{equation}
   \mathcal{A}_{\mu\nu}(x;\mathcal{D}_x)\equiv
   \frac{1}{2}\left[
   \varTheta_{\mu\nu}(x;\mathcal{D}_x)-\varTheta_{\nu\mu}(x;\mathcal{D}_x)
   \right].
\label{eq:(2.24)}
\end{equation}
The conservation of~$\mathcal{A}_{\mu\nu}(x;\mathcal{D}_x)$ can be shown by
the same argument as in~Ref.~\cite{Suzuki:2012gi}: Assuming the hypercubic
symmetry, it turns out that any dimension~$4$ anti-symmetric rank-$2$ tensor
can be expressed as\footnote{To apply this argument, the
operator~$\varTheta_{\mu\nu}(x;\mathcal{D}_x)$ must be local. This is actually
the case because, under any local variation of fields, the combination
$\sum_{y\in\mathcal{D}_x}\partial_\rho^S\mathcal{S}_\rho(y)$ is invariant.}
\begin{align}
   &\mathcal{A}_{\mu\nu}(x;\mathcal{D}_x)
\notag\\
   &=A_1\epsilon_{\mu\nu\rho\sigma}\partial_\rho^S
   \tr\left[\Bar{\psi}(x)\gamma_\sigma\gamma_5\psi(x)\right]
   +A_2\tr\left[\Bar{\psi}(x)\sigma_{\mu\nu}(D+M)\psi(x)\right]
   +a\mathcal{G}_{\mu\nu}(x),
\label{eq:(2.25)}
\end{align}
where $A_1$ and~$A_2$ are constants and the dimension~$5$
operator~$\mathcal{G}_{\mu\nu}(x)$ is at most logarithmically divergent. From
this general form, we have
\begin{equation}
   \left\langle
   \partial_\mu^S\mathcal{A}_{\mu\nu}(x;\mathcal{D}_x)
   \mathcal{O}
   \right\rangle
   \xrightarrow{a\to0}0,
   \qquad\text{for $x\leftrightsquigarrow\supp(\mathcal{O})$}.
\label{eq:(2.26)}
\end{equation}
This is trivially true for the first term in the right-hand side
of~Eq.~\eqref{eq:(2.25)}. For the second term in the right-hand side
of~Eq.~\eqref{eq:(2.25)}, this holds because of the equation of motion of the
gluino field. Finally, for the last term of~Eq.~\eqref{eq:(2.25)}, this follows
because of the overall factor of~$a$.

The combination of the above two properties, Eq.~\eqref{eq:(2.23)}
and Eq.~\eqref{eq:(2.26)} implies the conservation law of the symmetric part
of~$\varTheta_{\mu\nu}(x;\mathcal{D}_x)$, Eq.~\eqref{eq:(2.14)}, that is
\begin{equation}
   \left\langle
   \partial_\mu^S\mathcal{T}_{\mu\nu}(x;\mathcal{D}_x)
   \mathcal{O}
   \right\rangle
   \xrightarrow{a\to0}0,
   \qquad\text{for $x\leftrightsquigarrow\supp(\mathcal{O})$}.
\label{eq:(2.27)}
\end{equation}
This completes the proof of the conservation law of our lattice energy-momentum
tensor~\eqref{eq:(2.14)}.

For the new definition in Eqs.~\eqref{eq:(2.13)} and~\eqref{eq:(2.14)}, we can
further show that the expectation value of the energy density vanishes in the
continuum limit,
\begin{equation}
   \left\langle
   \mathcal{T}_{00}(x;\mathcal{D}_x)
   \right\rangle
   =\left\langle
   \varTheta_{00}(x;\mathcal{D}_x)
   \right\rangle
   \xrightarrow{a\to0}0,
\label{eq:(2.28)}
\end{equation}
when \emph{periodic boundary conditions\/} are imposed on all the fields. This
property of the energy density operator is natural from the perspective of
SUSY, because Eq.~\eqref{eq:(2.28)} corresponds to the derivative of the
supersymmetric partition function (i.e., the Witten index~\cite{Witten:1982df})
with respect to the temporal size of the system. In other words,
Eq.~\eqref{eq:(2.28)} shows that the origin of the energy that is consistent
with SUSY is \emph{automatically\/} chosen in the continuum limit; this is a
virtue of the present definition of the energy-momentum tensor compared with
our previous one~\cite{Suzuki:2012gi}.\footnote{For the definition of the
energy density operator in a lattice formulation of the two-dimensional
$\mathcal{N}=(2,2)$ SYM~\cite{Sugino:2003yb,Sugino:2004qd} (see also
Refs.~\cite{Cohen:2003xe,D'Adda:2005zk}) that possesses the
property~\eqref{eq:(2.28)} even before taking the $a\to0$ limit,
see~Refs.~\cite{Kanamori:2007ye,Kanamori:2007yx,Kanamori:2009dk,Kadoh:2009rw}.}
To show~Eq.~\eqref{eq:(2.28)}, we note that
$\sum_{y\in L^4}\partial_\rho^S\mathcal{S}_\rho(y)=0$ holds under the periodic
boundary conditions. From this,
\begin{align}
   \left\langle
   \varTheta_{00}(x;\mathcal{D}_x)
   \right\rangle
   &=\frac{1}{8}\left(C^{-1}\gamma_0\right)_{\alpha\beta}
   a^4\sum_{y\in L^4-\mathcal{D}_x}
   \left\langle
   \left[\partial_\rho^S\mathcal{S}_\rho(y)\right]_\alpha
   \left[\mathcal{S}_0(x)\right]_\beta
   \right\rangle
\notag\\
   &=\frac{1}{8}\left(C^{-1}\gamma_0\right)_{\alpha\beta}
   a^4\sum_{y\in L^4-\mathcal{D}_x}
   \left\langle\mathcal{Z}
   \left[a\mathcal{E}(y)\right]_\alpha
   \left[\mathcal{S}_0(x)\right]_\beta
   \right\rangle,
\label{eq:(2.29)}
\end{align}
where $L^4-\mathcal{D}_x$ denotes the complement of the region~$\mathcal{D}_x$
in the lattice~$L^4$ and we have used the SUSY WT relation~\eqref{eq:(2.1)} in
the second equality. Then since this is a correlation function of renormalized
operators with no mutual overlap with an overall factor of~$a$, this vanishes
in the continuum limit, i.e., Eq.~\eqref{eq:(2.28)} holds.

Our new definition in~Eqs.~\eqref{eq:(2.13)} and~\eqref{eq:(2.14)} contains
two unknown combinations of renormalization constants which must be determined
non-perturbatively. One is the overall normalization of~$\mathcal{S}_\mu(x)$,
$\mathcal{Z}\mathcal{Z}_S$ and other is the
ratio in~$\mathcal{S}_\mu(x)$, $\mathcal{Z}_T/\mathcal{Z}_S$. See
Eq.~\eqref{eq:(2.2)}. Among these, the latter
ratio~$\mathcal{Z}_T/\mathcal{Z}_S$ has been non-perturbatively measured in the
process to find the SUSY point in non-perturbative lattice simulations using
the Wilson fermion~\cite{Farchioni:2000mp,Farchioni:2001yn,Farchioni:2004fy,%
Demmouche:2010sf}. The former overall normalization~$\mathcal{Z}\mathcal{Z}_S$
may be determined from the expectation value of the energy
operator~$-a^3\sum_{\Vec{x}}\mathcal{T}_{00}(x;\mathcal{D}_x)$ in a certain
reference (e.g., one-particle) state. Thus, the determination of unknown
constants is much simpler than our previous construction
in~Ref.~\cite{Suzuki:2012gi} that requires the determination of other two
unknown constants, $\mathcal{Z}_{\text{EOM}}$ in~Eq.~\eqref{eq:(2.7)} and $c$
in~Eq.~\eqref{eq:(2.11)}. This point can be a great advantage in practical
applications.

On the other hand, the new definition has an $O(a)$ ambiguity associated with
the choice of the region~$\mathcal{D}_x$ in~Eq.~\eqref{eq:(2.13)} and this
ambiguity can be a possible source of the systematic error. Also, since the
energy-momentum tensor is defined by the product of two SUSY currents at
different points as~Eq.~\eqref{eq:(2.13)}, the application requires the
computation of correlation functions with the number of arguments as twice as
large compared with the correlation function of the energy-momentum tensor
(e.g., one defined in~Ref.~\cite{Suzuki:2012gi}). Only an implementation of the
present construction in actual numerical simulations will answer whether there
is a real payoff or not.

We believe that the basic idea on the construction of a lattice energy-momentum
tensor in the present Letter (and in~Ref.~\cite{Suzuki:2012gi}) is applicable
to more general 4D supersymmetric models. For our argument on the conservation
law of the renormalized SUSY current in the continuum limit to hold, however,
one has to carry out parameter fine tuning of sufficiently many numbers that
ensures the SUSY WT relation~\eqref{eq:(2.1)}. If such fine tuning is feasible
for the model under consideration, our idea to construct a lattice
energy-momentum tensor from the SUSY current will be useful to study physical
questions in supersymmetric models, such as the spontaneous SUSY breaking, the
mass and the decay constant of the pseudo Nambu--Goldstone boson associated
with the (classical) dilatation invariance and so on.

\section*{Acknowledgements}

I am indebted to Martin L\"uscher for discussions at the workshop ``New
Frontiers in Lattice Gauge Theory'' held at the Galileo Galilei Institute for
Theoretical Physics, Arcetri, Florence, of which the present work grew out. I
would like to thank also the workshop organizers for their great hospitality
and participants of the workshop for enjoyable discussions. This work is
supported in part by a Grant-in-Aid for Scientific Research, 22340069
and~23540330.





\bibliographystyle{elsarticle-num}
\bibliography{<your-bib-database>}



\end{document}